\documentclass[12pt]{article}
\usepackage[paper=letterpaper,hmargin=1.25in,top=1.25in,bottom=1.0in]{geometry}
\usepackage{amsfonts}

\newtheorem{theorem}{Theorem}
\newtheorem{lemma}{Lemma}

\newcounter{exmp}

\newenvironment{proof}[1][{}]{{\bf Proof{#1}.\ }}{\hfill\qed}
\newcommand{\binomial}[2]{\left(\begin{array}{c}{#1}\\{#2}\end{array}\right)}

\newcommand{\TT}[1][t]{{L}_{#1}}
\newcommand{\tinv}{t^{-1}}
\newcommand{\tnp}{[t,n,p]}
\newcommand{\tcp}{[t,\cdot,p]}

\newcommand{\filler}{{\scriptscriptstyle F}}
\newcommand{\tauf}{\tau_\filler}
\newcommand{\betaf}{\beta_\filler}
\newcommand{\nuf}{\nu_\filler}
\newcommand{\half}{\frac{1}{2}}
\newcommand{\entropy}{h}

\newcommand{\Triangle}{\triangle}

\newcommand{\qed}{\hspace*{\fill}%
    \vbox{\hrule\hbox{\vrule\squarebox{.667em}\vrule}\hrule}\smallskip}
    \def\squarebox#1{\hbox to #1{\hfill\vbox to #1{\vfill}}}

\title{\textbf{On the number of $t$-ary trees with a given path length}}%

\author{Gadiel Seroussi%
        \thanks{Hewlett-Packard Laboratories, Palo Alto, CA 94304,
        USA. Part of this work was done while the author was with
        the Mathematical Sciences Research Institute (MSRI), Berkeley,
        California, USA. E-mail: {\texttt{gseroussi@ieee.org}}. }%
}
\date{} 
\begin{document}
\maketitle

\begin{abstract}
\noindent
We show that the number of $t$-ary trees with path length equal to
$p$ is $\exp\big({\entropy(\tinv)t \log t\, {\frac{p}{\log p}
}\,(1+o(1))}\big)$, where $\entropy(x)={-}x\log  x {-}(1{-}x)\log
(1{-}x)\,$ is the binary entropy function. Besides its intrinsic
combinatorial interest, the question recently arose in the context
of information theory, where the number of $t$-ary trees with path
length $p$ estimates the number of universal types, or,
equivalently, the number of different possible Lempel-Ziv'78
dictionaries for sequences of length $p$ over an alphabet
of~size~$t$.

\vspace{6mm}

\noindent\textbf{Key words\ \ }{binary trees; $t$-ary trees; path length; universal types}
\end{abstract}

\section{Introduction}
Path length is an important global parameter of a tree that arises
in various computational contexts (cf.~\cite[Sec.\
2.3.4.5]{Knuth1}). Although the distribution of path lengths among
trees with a given number of nodes has been studied, the problem of
estimating their distribution by path length alone has remained
open. The question recently arose in an information-theoretic
context, in connection with the notion of
\emph{universal type}~\cite{Ser04LATIN,Ser04Utypes}, based on the
incremental parsing of Ziv and Lempel (LZ78)~\cite{LZ78}. When
applied to a $t$-ary sequence, the LZ78 parsing produces a
dictionary of strings that is best represented by a $t$-ary tree
whose path length corresponds to the length of the sequence. Two
sequences of the same length are said to be of the same universal
type if they yield the same $t$-ary parsing tree. Sequences of the
same universal type are, in a sense, statistically
indistinguishable, as the variational distance between their
empirical probability distributions of any finite order vanishes in
the limit~\cite{Ser04Utypes,Ser04LATIN}. Universal types generalize
the notion underlying the classical
\emph{method of types}, which has lead to important theoretical
results in information theory~\cite{Csi98}. Of great interest in
this context is the estimation of the number of different types for
sequences of a given length. For universal types, this translates to
the number of different LZ78 dictionaries for $t$-ary sequences of a
given length, or, equivalently, the number of $t$-ary trees with a
given path length, which is the subject of this paper.

First, we present some definitions and formalize the problem. Fix an
integer $t\geq 2$. A $t$-ary tree $T$ is defined recursively as
either being empty or consisting of a
\emph{root node} $r$ and the nodes of $t$ disjoint, ordered, $t$-ary
(sub-) trees $T_1,T_2,\ldots,T_t$, any number of which may be empty
(cf.~\cite[Sec.\ 2.3.4.5]{Knuth1}). When $T_i$ is not empty, we say
that there is an \emph{edge} from $r$ to the root $r'$ of $T_i$, and
that $r'$ is a \emph{child} of $r$. The total number of nodes of $T$
is zero if $T$ is empty, or $n_T=1+\sum_{i=1}^t n_{T_i}$ otherwise.
A node of $T$ is called a
\emph{leaf} if it has no children. The
\emph{depth} of a node $v\in T$ is defined as the number of
edges traversed to get from the root $r$ to $v$. We denote by
$D_{j}^{(T)}$, $j\geq 0$, the number of nodes at depth $j$ in $T$.
The sequence $\{D_j^{(T)}\}$ is called the
\emph{profile} of $T$; we consider only finite trees, so $\{D_j^{(T)}\}$
has finite support. The
\emph{path length} of a non-empty tree $T$, denoted by $p_T$, is the sum of the
depths of all the nodes in $T$, namely
\[
p_T =\sum_{j\geq 1} j D_{j}^{(T)}
\]
The subscript $T$ in $n_T$ and $p_T$ will be omitted in the sequel
when the tree being discussed is clear from the context. We call a
$t$-ary tree with $n$ nodes a $[t,n]$
\emph{tree}. A $[t,n]$ tree with path length
equal to $p$ will be called a $\tnp$ \emph{tree}, and a $t$-ary
tree with path length equal to $p$ and an unspecified number of
nodes will be referred to as a $\tcp$ \emph{tree}.

Let $C_t(n)$ denote the number of $[t,n]$ trees, and $\TT(p)$ the
number of $\tcp$ trees. It is well known~\cite[p.\ 589]{Knuth1} that
\begin{equation}\label{eq:catalan-t}
C_t(n) = \frac{1}{(t-1)n+1}\binomial{tn}{n}\,,\quad n\geq 0,\,t>1.
\end{equation}
In the binary case ($t=2$), these are the well known \emph{Catalan
numbers} that arise in many combinatorial contexts. The
determination of $\TT(p)$, on the other hand, has remained
elusive, even for $t=2$. Consider the bivariate generating
function $B(w,z)$ defined so that the coefficient of $w^pz^n$ in
$B(w,z)$ counts the number of $[2,n,p]$ trees. $B(w,z)$ satisfies
the functional equation~\cite[p.\ 595]{Knuth1}
\[zB(w,wz)^2 = B(w,z) - 1.\]
However, deriving the generating function, $B(w,1)$, of the numbers
$\TT[2](p)$ from this equation appears quite challenging.
Nevertheless, the equation and others of similar structure have been
studied in the literature. In particular, the limiting distribution
of the path length for a given number of nodes is related to the
area under a Brownian
excursion~\cite{Takacs91a,Takacs91b,Takacs91c}, which is also known
as an
\emph{Airy distribution}. This distribution occurs in many
combinatorial problems of theoretical and practical interest
(cf.~\cite{FlajoLouch01} and references therein). These studies,
however, have not yielded explicit asymptotic estimates for the
numbers $\TT(p)$.

Let $\entropy(x)=-x\log x {-}(1{-}x)\log(1{-}x)$ denote the binary
entropy function.\footnote{
\nopagebreak{Unless a base is explicitly specified,
$\exp$ and $\log$ denote, respectively, the exponential and
logarithm functions with respect to an arbitrary base that remains
consistent throughout the paper.}}%
The main result of this paper is the following asymptotic estimate
of $\TT(p)$.
\begin{theorem}\label{th:main}
Let $\alpha=\entropy(\tinv)\,t\log t$. Then,
\(
\TT(p) = \exp\Big(\,{{\displaystyle\frac{\alpha p}{\log p} }\big(1+o(1)\big)}\,\Big).
\)
\end{theorem}
The theorem is derived by proving matching upper and lower bounds
on $\log\TT(p)$. The proof is presented in
Section~\ref{sec:proof}.

We remark that Knessl and Szpankowski~\cite{KnesslSzpa04} have
recently applied the WKB heuristic~\cite{WKB} to obtain an
asymptotic expansion of $\log \TT[2](p)$ using tools of complex
analysis. The heuristic makes certain assumptions on the form of
asymptotic expansions, and is often considered a practically
effective albeit non-rigorous method.  The main term in the
expansion of~\cite{KnesslSzpa04} is consistent with
Theorem~\ref{th:main} for $t=2$. The proofs in this paper, presented
in the next section, are based mostly on simple combinatorial
arguments.

\section{Proof of the main result}\label{sec:proof}
In the following lemma, we list some elementary properties of
$t$-ary trees that will be referred to in the proof of
Theorem~\ref{th:main}. For a discussion of these properties,
see~\cite[Sec.~2.3.4.5]{Knuth1}.\footnote{A slight change of
terminology is required: nodes of $t$-ary trees in our terminology
correspond to
\emph{internal} nodes of \emph{extended} $t$-ary trees in~\cite{Knuth1}.}

\begin{lemma}\label{lem:aux}
\begin{description}
\item{(i)}
Let $\ell$ be a positive integer, and let $T$ be a $[t,n,p]$ tree
achieving minimal path length among all $t$-ary trees with $\ell$
leaves. Then,
\begin{equation}\label{eq:l}
n =\ell + \left\lceil \frac{\ell-1}{t-1}\right\rceil \,.
\end{equation}
Define
\begin{equation}\label{eq:m} m = \lceil \,\log_t \ell \,\rceil,
\end{equation}
and
\begin{equation} \label{eq:l1}
\ell_1 = \ell - \left\lfloor\frac{t^m-\ell}{t-1}\right\rfloor\,.
\end{equation}
Then, the profile of $T$ is given by
\begin{equation}\label{eq:profile}
D_j^{(T)} = \left\{\;\begin{array}{ll} %
t^j,& 0 \leq j \leq  m-1,\\
\ell_1, & j = m,\\
0,      & j > m,
\end{array}\right.
\end{equation}
In particular, all the leaves of $T$ are either at depth $m$ or
$m-1$.
\item{(ii)}
A $[t,n,p]$ tree with minimal path length satisfies
\begin{equation}\label{eq:pmin}
p = p_{\min} = \left(n + \frac{1}{t-1}\right) \mu -
\frac{t(t^{\mu}-1)}{(t-1)^2} = n \log_t n - O(n),
\end{equation}
where $\mu = m$ whenever $n \not\equiv 2 \bmod t$, or $\mu =
m{+}1$ otherwise, with $m$ defined in~(\ref{eq:m}) for the number
of leaves, $\ell$, of the tree. In particular, the tree of (i)
satisfies~(\ref{eq:pmin}) with $\mu=m$.
\item{(iii)}
The number of nodes of a $[t,n,p]$ tree satisfies
\begin{equation}\label{eq:nmax}
n \leq \frac{p}{\log_t p - O(\log\log p)} = \frac{p}{\log_t
p}(1+o(1)).
\end{equation}
\item{(iv)} The maximal path length of a $[t,n]$ tree is achieved
by a tree in which each internal node has exactly one child (hence,
there is exactly one leaf in the tree). The path length of such a
tree is
\begin{equation}\label{eq:pmax}
p_{\max} = \frac{n(n-1)}{2}\;.
\end{equation}
\item{(v)} There is a $[t,n,p]$ tree for each $p$ in the range
$p_{\min}\leq p \leq p_{\max}$.
\end{description}
\end{lemma}
\begin{proof}
Items (i),(ii), and (iv) follow immediately from the discussion
in~\cite[Sec.~2.3.4.5]{Knuth1}. For convenience in the proof of
Theorem~\ref{th:main}, we characterize, in Item (i), trees with
minimal path length for a given number of
\emph{leaves}, while the discussion in~\cite{Knuth1} does so for
trees with a given number of
\emph{nodes}. The two characterizations coincide, except
for values of $n$ such that $n \equiv\ 2 \bmod t$, which never
occur in~(\ref{eq:l}). In that case, a tree with $n-1$ nodes would
have the same number of leaves and a shorter path length. A tree
that has minimal path length for its number of leaves, on the
other hand, always has minimal path length also for its number of
nodes (given in~(\ref{eq:l})).

Item (iii) follows from (ii) by solving for $n$ in an equation of
the form $p = n\log_t n - O(n)$. Solutions of equations of this
form are related to the
\emph{Lambert W} function, a detailed discussion of which can be found
in~\cite{Lambert-function}.

To prove the claim of Item (v), consider a $[t,n,p]$ tree $T$ such
that $D_j^{(T)}>1$ for some integer $j$. Let $j_{T}$ be the
largest such integer for the tree $T$. It follows from these
assumptions that $T$ must have nodes $u$ and $v$ at depth $j_{T}$,
such that $u$ is a leaf, $v\neq u$, and $v$ has at most one child.
Thus, we can transform $T$ by deleting $u$ and adding a child to
$v$, and obtain a $[t,n,p+1]$ tree. Starting with a
$[t,n,p_{\min}]$ tree, the transformation can be applied
repeatedly to obtain a sequence of trees
 with consecutive values of $p$, as long as the transformed tree has at
least two leaves. When this condition ceases to hold, we have the
tree of Item (iv), which has path length $p_{\max}$.
\end{proof}

We will also rely on the following estimate of $C_t(n)$ derived
from~(\ref{eq:catalan-t}) using Stirling's approximation (see,
e.g.,~\cite[Ch.\ 10]{MacWillSloane:83}). For positive real numbers
$c_1$ and $c_2$, which depend on $t$ but not on $n$, we have
\begin{equation}\label{eq:Capprox}
c_1 n^{-\frac{3}{2}} \exp\big(h(\tinv)t\, n \big)\leq C_t(n) \leq
c_2 n^{-\frac{3}{2}} \exp\big(h(\tinv)t\, n \big).
\end{equation}

\noindent\textbf{Proof of Theorem~\ref{th:main}.}
We recall that $\alpha=\entropy(\tinv)\,t\log t$.

\textbf{(a) Upper bound:}
$\log \TT(p) \leq {{\displaystyle\frac{\alpha p}{\log p}
}(1+o(1))}$.

Let $n_p$ denote the maximum number of nodes of any tree with path
length equal to $p$. Clearly, we have
\[
\TT(p) \leq \sum_{n=1}^{n_p} C_t(n) \leq n_p C_t(n_p)
\]
and thus, by~(\ref{eq:Capprox}), we obtain
\begin{eqnarray}\label{eq:ub1}
\log \TT(p) &\leq& \log n_p + \log C_t(n_p) \leq h(\tinv)t\, n_p
-\frac{1}{2}\log n_p + O(1)\nonumber\\
&=& \frac{\alpha}{\log t} n_p -
\frac{1}{2} \log n_p + O(1).
\end{eqnarray}
The claimed upper bound on $\log\TT(p)$ follows
from~(\ref{eq:ub1}) by applying Lemma~\ref{lem:aux}(iii) with
$n{=}n_p$. The asymptotic error term $o(1)$ in the upper bound is,
by~(\ref{eq:nmax}), of the form $O(\log\log p/\log p)$.

\textbf{(b) Lower bound:}
$\log \TT(p) \geq {{\displaystyle\frac{\alpha p}{\log p}
}(1+o(1))}$.

We prove the lower bound by constructing a sufficiently large
class of $\tcp$ trees.

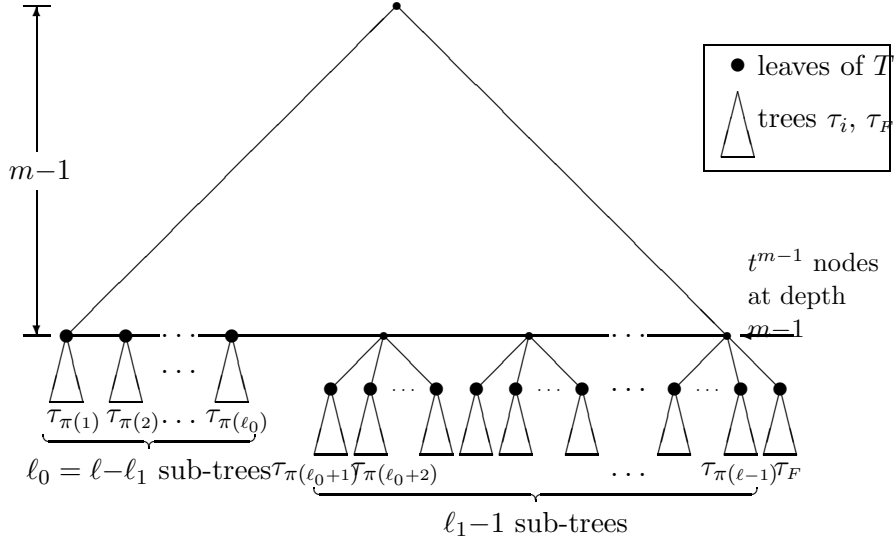
\begin{figure}
\begin{center}
\setlength{\unitlength}{2.5pt}
\newsavebox{\littletreebare}
\savebox{\littletreebare}{
\put(0,0){\line(-1,-4){2.5}}
\put(0,0){\line(1,-4){2.5}}
\put(-2.5,-10){\line(1,0){5}}
}
\newsavebox{\littletree}
\savebox{\littletree}{
\put(0,0){\circle*{2}}
\put(0,0){\usebox{\littletreebare}}
}

\newsavebox{\children}
\savebox{\children}{
\put(0,0){\circle*{1}}
\put(0,0){\line(-1,-1){8}}
\put(0,0){\line(-1,-4){2}}
\put(0,0){\line(1,-1){8}}
\put(0,-8.4){\makebox(6,0)[b]{\tiny$\ldots$}}
}
\newsavebox{\revchildren}
\savebox{\revchildren}{
\put(0,0){\circle*{1}}
\put(0,0){\line(-1,-1){8}}
\put(0,0){\line(1,-4){2}}
\put(0,0){\line(1,-1){8}}
\put(-6,-8.4){\makebox(6,0)[b]{\tiny$\ldots$}}
}
\newsavebox{\childrenwithtrees}
\savebox{\childrenwithtrees}{
\put(0,0){\usebox{\children}}
\put(-8,-8){\usebox{\littletree}}
\put(-2,-8){\usebox{\littletree}}
\put(+8,-8){\usebox{\littletree}}
}
\newsavebox{\revchildrenwithtrees}
\savebox{\revchildrenwithtrees}{
\put(0,0){\usebox{\revchildren}}
\put(-8,-8){\usebox{\littletree}}
\put( 2,-8){\usebox{\littletree}}
\put(+8,-8){\usebox{\littletree}}
}
\newsavebox{\childrenwithtworighttrees}
\savebox{\childrenwithtworighttrees}{
\put(0,0){\usebox{\children}}
\put(-2,-8){\usebox{\littletree}}
\put(+8,-8){\usebox{\littletree}}
}
\newsavebox{\childrenwithnodes}
\savebox{\childrenwithnodes}{
\put(0,0){\usebox{\children}}
\put(-8,-8){\circle*{1}}
\put(-2,-8){\circle*{1}}
\put(+8,-8){\circle*{1}}
}
\newsavebox{\rightchild}
\savebox{\rightchild}{
\put(0,0){\circle*{1}}
\put(0,0){\line(1,-1){8}}
}
\newsavebox{\rightchildwithtree}
\savebox{\rightchildwithtree}{
\put(0,0){\usebox{\rightchild}}
\put(+8,-8){\usebox{\littletree}}
}
\newsavebox{\rightchildwithnode}
\savebox{\rightchildwithnode}{
\put(0,0){\usebox{\rightchild}}
\put(+8,-8){\circle*{1}}
}
\newcommand{\braces}[2]{
    \put(-1,00){\oval(#1,2)[lb]}
    \put(-1,-2){\oval(02,2)[tr]}
    \put(01,-2){\oval(02,2)[tl]}
    \put(01,00){\oval(#1,2)[rb]}
    \put(00,-5.5){\makebox(0,0){#2}}}

\begin{picture}(120,80)(0,-30)
\put(0,0){\line(1,1){50}}
\put(50,50){\line(1,-1){50}}
\put(0,0){\line(1,0){14}}
\put(14,0){\makebox(6,0){$\ldots$}}
\put(20,0){\line(1,0){62}}
\put(82,0){\makebox(6,0){$\ldots$}}
\put(88,0){\line(1,0){12}}
\put(50,50){\circle*{1}}

\put(0,0){
\put(00,0){\usebox{\littletree}}
\put(-3,-13){\small$\tau_{\scriptstyle\pi({\scriptscriptstyle 1})}$}
\put(09,0){\usebox{\littletree}}
\put(6,-13){\small$\tau_{\scriptstyle\pi({\scriptscriptstyle 2})}$}
\put(17,-5.4){\makebox(0,0){$\ldots$}}
\put(25,0){\usebox{\littletree}}
\put(17,-13){\makebox(0,0){$\ldots$}}
\put(21,-13){\small$\tau_{\scriptstyle\pi({\scriptscriptstyle\ell_0})}$}
\put(12.5,-15){\braces{30}{{\small $\ell_0=\ell{-}\ell_1$ sub-trees}}}
}

\put(48,0) {
\put(0,0){\usebox{\childrenwithtrees}}

\put(-8,-8){\circle*{1}}
\put(-17,-21){\small$\tau_{\scriptstyle\pi({\scriptscriptstyle\ell_0{+}1})}$}
\put(-5,-21){\small$\tau_{\scriptstyle\pi({\scriptscriptstyle\ell_0{+}2})}$}

\put(22,0){\usebox{\childrenwithtrees}}
\put(34,-8.4){\makebox(6,0)[b]{$\ldots$}}
\put(52,0){\usebox{\revchildrenwithtrees}}
\put(34,-21){\makebox(6,0){$\ldots$}}
\put(48,-21){\small$\tau_{\scriptstyle\pi({\scriptscriptstyle\ell{-}1})}$}
\put(59,-21){\small$\tauf$}
\put(23,-23){\braces{65}{{\small $\ell_1{-}1$ sub-trees}}}
}

\put(110,0){\vector(-1,0){8}}
\put(108,1){\makebox(10,10){
    $\begin{array}{l}\mbox{\footnotesize $t^{m-1}$ nodes}\\[-2pt]
                   \mbox{\footnotesize at depth}\\[-2pt]
                   \mbox{\footnotesize $m{-}1$}\end{array}$
}}
\put(0.5,0){
\put(-5,22){\vector(0,-1){22}}
\put(-7,0){\line(1,0){4}}
\put(-5,28){\vector(0,1){22}}
\put(-7,50){\line(1,0){4}}
\put(-7,25){\makebox(5,0){\small$m{-}1$}}
}
\put(95,25) {
\framebox(28,19)[l]{
  \put(5,-1){
    \put(0,0){\circle*{2}}
    \put(3,0){\makebox(0,0)[l]{\small leaves of $T$}}
    \put(0,-4){\usebox{\littletreebare}}
    \put(3,-6){\makebox(0,-5)[l]{\small trees $\tau_i,\,\tauf$}}
}} }

\end{picture}
\normalsize
\end{center}
\caption{\label{fig:Tpi}Tree $T_\pi$}
\end{figure}

Let $\ell$ be a positive integer, $\ell>2$. We start with a $t$-ary
tree $T$ with $\ell$ leaves and shortest possible path length, as
characterized in Lemma~\ref{lem:aux}(i). Let $q$ be the integer
satisfying
\begin{equation}\label{eq:Ctq}
C_t(q-1) <\ell-1 \leq C_t(q)\,,
\end{equation}
and let $\tau_1,\tau_2,\ldots,\tau_{\ell-1}$ be the first $\ell{-}1$
distinct $[t,q]$ trees, when $[t,q]$ trees are arranged in
non-decreasing order of path length. Additionally, let $\tauf$ be a
tree with $\betaf q$ nodes, for some positive constant $\betaf$ to
be specified later. Finally, let $\pi$ be a permutation on
$\{1,2,\ldots,\ell-1\}$. We construct a tree $T_\pi$ by attaching
the trees $\tau_1,\tau_2,\ldots,\tau_{\ell-1}$ and $\tauf$ to the
leaves of $T$, so that the $i$-th leaf (taken in some fixed order)
becomes the root of a copy of $\tau_{\pi(i)}$, $1\leq i <
\ell$. The tree $\tauf$, in turn, is attached to the last leaf of $T$,
which is assumed to be at (the maximal) depth $m$. The
construction is illustrated in Figure~\ref{fig:Tpi}.

\normalsize

Next, we compute the path length, $p$,  of $T_\pi$. By
Lemma~\ref{lem:aux}(i), all the leaves of $T$ are either at depth
$m=\lceil \log_t \ell\rceil$ or at depth $m{-}1$. Assume $\tau_i$,
$1\leq i \leq
\ell{-}1$, is attached to a leaf of depth $m{-}1{+}\epsilon_i$,
$\epsilon_i\in
\{0,1\}$, of $T$. Also, let $\nu_i$ denote the path length of $\tau_i$,
$1\leq i \leq \ell{-}1$, and $\nuf$ the path length of $\tauf$. The
contribution of $\tau_i$ (excluding its root) to $p$ is
\begin{eqnarray*}
p_i &=& \sum_{j\geq 1} (m-1+\epsilon_i+j)D_j^{(\tau_i)} =
(m-1+\epsilon_i)\sum_{j\geq 1}D_j^{(\tau_i)} +\sum_{j\geq 1}j
D_j^{(\tau_i)} \\
&=& (m-1+\epsilon_i)(q-1) + \nu_i,
\end{eqnarray*}
Similarly, the contribution of $\tauf$ to $p$ is
\[ p_\filler = m(\betaf q-1) +
\nuf\,.
\]
Considering also the contribution of $T$ according to its
profile~(\ref{eq:profile}), we obtain
\begin{equation}
p =  \sum_{i=1}^{\ell-1} (m-1+\epsilon_i)(q-1)+\sum_{i=1}^{\ell-1}
\nu_i + m(\betaf q - 1) +
\nuf +
\sum_{j=1}^{m-1}j t^j
+ \ell_1 m\,.
\label{eq:p0}
\end{equation}
Further, observing that $\sum_{i=1}^{\ell-1}\epsilon_i = \ell_1$,
and defining $\overline{\nu}
=(\ell-1)^{-1}\sum_{i=1}^{\ell-1}\nu_i$, we obtain
\begin{equation}
p = \left((\ell-1)(m-1)+\ell_1\right)(q-1) +(\ell-1)\overline{\nu}
+
 m (\betaf q-1) + \nuf + \sum_{j=1}^{m-1} j t^j + \ell_1 m\,.
\label{eq:p1}
\end{equation}
Recall that the trees $\tau_i$ were selected preferring shorter path
lengths, so their average path length $\overline{\nu}$ is at most as
large as the average path length of
\emph{all} $[t,q]$ trees. The latter average is known to be $O(q^{3/2})$ (this
follows from the results of~\cite{FlajOdlyz82}; see also~\cite[Sec.\
2.3.4.5]{Knuth1} for $t=2$). Observe also that, from the definition
of $q$ in~(\ref{eq:Ctq}), using~(\ref{eq:Capprox}) and (\ref{eq:m}),
and recalling that $\alpha=h(\tinv)t\log t$, we obtain
\begin{equation}\label{eq:q(m)}
q = \frac{\log^2 t}{\alpha} m + O(\log m).
\end{equation}
Recalling now that $n_{\tau_{\scriptscriptstyle F}}=\betaf q$, and,
hence, $\nuf = O(q^2)$, it follows, after standard algebraic
manipulations, that~(\ref{eq:p1}) can be rewritten as
\begin{equation}\label{eq:p2}
p =\frac{\log^2 t}{\alpha}\, m^2\ell + O(m^{3/2}\ell).
\end{equation}
It also follows from~(\ref{eq:p1}) that $p$ is independent of the
choice of permutation $\pi$. Moreover, by construction, each
permutation $\pi$ defines a different tree $T_\pi$, and, therefore,
we have
\begin{equation}
\label{eq:ell!}\TT(p)\geq (\ell-1)!\,.
\end{equation}
From~(\ref{eq:ell!}), using Stirling's approximation,
applying~(\ref{eq:m}) and~(\ref{eq:p2}), and simplifying, we can
write
\begin{eqnarray}
\frac{\log \TT(p)}{p} &\geq&
\frac{\log(\ell-1)!}{p} = \frac{\ell\log\ell -
O(\ell)}{p}\nonumber\\
& =&\frac{\ell m \log t  - O(\ell)}{\alpha^{-1}\, (\log^2
t)\,m^2\ell + O(m^{3/2}\ell)} =
\frac{\alpha\left(1-O(m^{-1})\right)}{m\log t\left(1+O(m^{-\half})\right)}\;.
\label{eq:lbound0}
\end{eqnarray}
Taking logarithms on both sides of~(\ref{eq:p2}), and
applying~(\ref{eq:m}), we can write $m\log t  =
\log p - O(\log m)$. Substituting for $m\log t$ in~(\ref{eq:lbound0}), and
simplifying asymptotic expressions, we obtain
\begin{equation}
\frac{\log  \TT(p)}{p} \geq
\frac{\alpha}{\log p}(1-o(1)),
\label{eq:lbound}
\end{equation}
from which the desired lower bound follows. The $o(1)$ term
in~(\ref{eq:lbound}) is $O(m^{-\half})=O((\log p)^{-\half})$.

The above construction yields large classes of trees of path length
$p$ for a sparse sequence of values of $p$, controlled by the
parameter $\ell$. Next, we show how the gaps in the sparse sequence
can be filled, yielding constructions, and validating the lower
bound, for all (sufficiently large) integer values of $p$. In the
following discussion, when we wish to emphasize the dependency of
$m$, $\ell_1$, $q$, and $p$ on $\ell$, we will use the notations
$m(\ell),\ell_1(\ell), q(\ell)$, and $p(\ell)$, respectively. Also,
for any such function $f(\ell)$, we denote by $\Triangle f$ the
difference $f(\ell+1)-f(\ell)$, with the value of $\ell$ being
implied by the context. We start by estimating $\Triangle p$.

Assume first that $\ell$ is such that $\Triangle q = 0$ and
$\Triangle m = 0$. Then, substituting $\ell+1$ for $\ell$
in~(\ref{eq:p1}), and subtracting the original equation, we obtain
\begin{equation}\label{eq:Deltap}
\Triangle p = (m-1 + \Triangle\ell_1)(q-1) +
\nu_\ell +\Triangle \ell_1\, m.
\end{equation}
It follows from~(\ref{eq:l1}) that, with $m$ fixed, we have $0\leq
\Triangle\ell_1 \leq 2$. Also, by~(\ref{eq:pmax}), we have $\nu_\ell <
\half q^2$. Hence, recalling~(\ref{eq:q(m)}), it follows
from~(\ref{eq:Deltap}) that
\begin{equation}\label{eq:Deltap2}
\Triangle p < \left(\frac{\alpha}{\log^2 t} + \half\right)q^2 + O(q\log
q).
\end{equation}
Notice that, in~(\ref{eq:p1}), with all other parameters of the
construction staying fixed, any increment in $\nuf$ produces an
identical change in $p$. By Lemma~\ref{lem:aux}(v), by an
appropriate evolution of $\tauf$, we can make  $\nuf$ assume any
value in the range $(\nuf)_{\min}\leq \nuf \leq (\nuf)_{\max}$,
where $(\nuf)_{\min}= O(\betaf q
\log q)$, and $(\nuf)_{\max} = \half \betaf q ( \betaf q - 1)$.
Choosing $\betaf > \sqrt{2\alpha (\lg t)^{-2} + 1}$, this range of
$\nuf$ will make $p$ span the gap between $p(\ell)$ and $p(\ell+1)$
as estimated in~(\ref{eq:Deltap2}), for all sufficiently large
$\ell$ satisfying the conditions of this case. Still, the variation
in the value of $p$ is asymptotically negligible and does not affect
the validity of~(\ref{eq:lbound}).

If $\Triangle m = 1$, we must have $\ell = \ell_1(\ell) = t^m$, and
$\ell_1(\ell+1) = 2$. In this case, using~(\ref{eq:p1}) again, we
obtain
\begin{eqnarray*}
\Triangle p &=& (\ell m + 2)(q-1) + \betaf q{-}1 + \nu_\ell + m \ell + 2(m+1)\\
& &  -
((\ell-1)(m-1)+\ell)(q-1) - \ell m \\
&=& (m+1)(q+1) + \nu_\ell + \betaf q - 1,
\end{eqnarray*}
which admits the same asymptotic upper bound as $\Triangle p$
in~(\ref{eq:Deltap2}). Thus, the gap between $p(\ell)$ and
$p(\ell+1)$ is filled also in this case by tuning the structure of
$\tauf$.

The above method cannot be applied directly when $\Triangle q = 1$.
We call a value of $\ell$ such that $q(\ell+1)=q(\ell)+1$ a
$q$-\emph{break}. At a $q$-break, $\Triangle p$ is exponential in
$q$, and a tree $\tauf$ of polynomial size cannot compensate for
such a gap. However, we observe that the construction of $T_\pi$,
and its asymptotic analysis in~(\ref{eq:p1})--(\ref{eq:lbound})
would also be valid if we chose $q' = q+1$, instead of $q$, as the
size of the trees $\tau_i$. This choice would produce a different
sequence of path length values $p'(\ell)$, which, when substituted
for $p$, would also satisfy~(\ref{eq:lbound}) and would validate the
lower bound of the theorem. It follows from~(\ref{eq:p1}) that
$p'(\ell) > p(\ell)$. Equivalently, for any given (sufficiently
large) value $\ell$, there exists an integer $\ell'<\ell$ such that
$p'(\ell')
\leq p(\ell) \leq p'(\ell'+1)$.

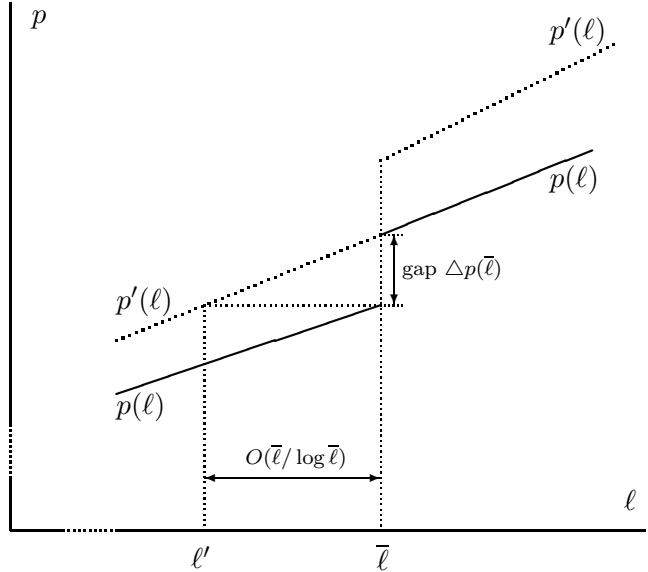
\begin{figure}
\begin{center}
\setlength{\unitlength}{4pt}

\begin{picture}(50,60)(-10,-10)
\thicklines
\put(0,0){\qbezier[37](25,30)(36,35.5)(47,41)} 
\put(0,0){\qbezier[40](00,13)(12.5,18)(25,23)} 
\put(25,23){\line(5,2){20}}
\put(0,8){\line(3,1){25}}
\thinlines

\put(8.333,-5){\qbezier[42](0,21.333)(0,10.667)(0,0)}
\put(25,-5){\qbezier[70](0,0)(0,17.5)(0,35)}
\put(25,16.333){\qbezier[5](0,0)(1,0)(2,0)}
\put(25,23){\qbezier[5](0,0)(1,0)(2,0)}
\put(25,16.333){\qbezier[32](0,0)(-8.333,0)(-16.667,0)}

\put(26.25,19){\vector(0,1){4}}
\put(26.25,19){\vector(0,-1){2.666}}
\put(27,19.667){\makebox(0,.5)[l]{{\scriptsize gap }{\scriptsize $\Triangle
p(\overline{\ell})$}}}

\put(8,-7.5){\makebox(0,0){\small$\ell'$}}
\put(25.2,-7.5){\makebox(0,0){\small$\overline{\ell}$}}
\put(0,6){{\small$p(\ell)$}}
\put(0,16){{\small$p'(\ell)$}}
\put(41,27.5){{\small$p(\ell)$}}
\put(41,41.5){{\small$p'(\ell)$}}


\thicklines
\put(-10,-5){\line(1,0){05}}
\put(-05,-5){\qbezier[10](0,0)(2.5,0)(5,0)}
\put(00,-05){\line(1,0){50}}
\put(48,-03){{\small$\ell$}}

\put(-10,-5){\line(0,1){05}}
\put(-10,0){\qbezier[10](0,0)(0,2.5)(0,5)}
\put(-10,05){\line(0,1){40}}
\put(-8,43){{\small$p$}}
\thinlines

\put(16.333,0){\vector(-1,0){8}}
\put(15,0){\vector(1,0){10}}
\put(17,2){\makebox(0,0){{\scriptsize$O(\overline{\ell}/\log\overline{\ell})$}}}
\end{picture}
\normalsize
\end{center}
\caption{\label{fig:qgap}Bridging the gap in $q$-breaks}
\end{figure}
\normalsize

Consider a $q$-break $\overline{\ell}$. To construct large classes
of trees for all values of $p$, proceed as follows (refer to
Figure~\ref{fig:qgap}): use the original sequence of values
$p(\ell)$, filling the gaps as described above, until $\ell =
\overline{\ell}$. At that point, find the largest integer $\ell'$
such that $p'(\ell') \leq p(\overline{\ell})$, and ``backtrack''
to $\ell=\ell'$. Continue with the sequence $p'(\ell)$, $\ell =
\ell',\ell'+1,
\ldots$, filling the gaps accordingly. Notice that $q'(\ell)$ to the left
of $\overline{\ell}$ is the same as $q(\ell)$ to the right of that
point. Thus, $p'(\ell)$ continues ``smoothly'' (i.e., with gaps
$\Triangle p$ as in~(\ref{eq:Deltap2})) into $p(\ell)$ at
$\ell=\overline{\ell}$.  The process now rejoins the sequence
$p(\ell)$ as before, until the next $q$-break point.
By~(\ref{eq:p1}), since the function $m(\ell)$ remains the same for
both $p$ and $p'$, we have, asymptotically,
\[\ell'
\approx (1-1/q)\overline{\ell} \approx \overline{\ell} -
c_3\overline{\ell}/\log\overline{\ell}\,,
\]
for some positive constant $c_3$. Thus, for sufficiently large
$\overline{\ell}$, although the difference between $\ell'$ and
$\overline{\ell}$ is negligible with respect to $\overline{\ell}$,
$\ell'$ is guaranteed to fall properly between $q$-breaks, and the
number of sequence points $p'(\ell)$ used between $\ell'$ and
$\overline{\ell}$ is unbounded.\qed

\noindent{\textbf{Acknowledgment.} Thanks to
Wojciech Spankowski and  Alfredo Viola for very useful discussions.
Also, the stimulating environment of the Tenth Analysis of
Algorithms seminar at MSRI in June of 2004 provided inspiration that
helped pin down the final details of the proof of Theorem 1.}

\bibliographystyle{siam}
\end{document}